\documentclass[aps,pre,twocolumn,groupedaddress,showpacs,showkeys,floatfix]{revtex4-1}
\usepackage{wrapfig}
\usepackage{graphicx}
\usepackage{colordvi}
\usepackage{color}
\usepackage{amsmath}
\usepackage{amssymb}
 \usepackage{comment}
\usepackage{natbib}

\usepackage[small, bf]{caption}

\newcommand{\ve}[1]{\mbox{\boldmath $#1$}}

\begin{document}
\title{Motif Analysis for Small-Number Effects in Chemical Reaction Dynamics}
%\shorttitle: Motif Analysis for Small-Number Effects

\author{Nen Saito}
\affiliation{
Research Center for Complex Systems Biology,
Graduate School of Arts and Sciences,
The University of Tokyo, 3-8-1 Komaba, Meguro-ku, Tokyo 153-8902, Japan }
\author{Yuki Sughiyama}
\affiliation{Institute of Industrial Science, The University of Tokyo,
4-6-1, Komaba, Meguro-ku, Tokyo 153-8505 Japan}
\author{Kunihiko Kaneko}\affiliation{
Research Center for Complex Systems Biology,
Graduate School of Arts and Sciences,
The University of Tokyo, 3-8-1 Komaba, Meguro-ku, Tokyo 153-8902, Japan}

\begin{abstract}
The number of molecules involved in a cell or subcellular structure is sometimes rather small. In this situation, ordinary macroscopic-level fluctuations can be overwhelmed by non-negligible large fluctuations, which results in drastic changes in chemical-reaction dynamics and statistics compared to those observed under a macroscopic system (i.e., with a large number of molecules). In order to understand how salient changes emerge from fluctuations in molecular number, we here quantitatively define small-number effect by focusing on a `mesoscopic' level, in which the concentration distribution is distinguishable both from micro- and macroscopic ones, and propose a criterion for determining whether or not such an effect can emerge in a given chemical reaction network. Using the proposed criterion, we systematically derive a list of motifs of chemical reaction networks that can show small-number effects, which includes motifs showing emergence of the power law and the bimodal distribution observable in a mesoscopic regime with respect to molecule number. The list of motifs provided herein is helpful in the search for candidates of biochemical reactions with a small-number effect for possible biological functions, as well as for designing a reaction system whose behavior can change drastically depending on molecule number, rather than concentration.

\end{abstract}
\maketitle

A living cell consists of a wide variety of biomolecules, which are encapsulated within a small cellular volume.
As an inevitable consequence, some chemical species have a small number of molecules.
In fact, recent advances in single-molecule measurement techniques have revealed that the copy numbers of different proteins in a living cell range widely; specifically, some protein species exist in quite low numbers in a bacterial cell~\cite{taniguchi2010quantifying,ishihama2008protein}.
Even in the case of eukaryotic cells, protein abundance in a fine intra-cellular
structure (such as the dendritic spine or organelle) is expected to be quite small.
This small molecule number leads to non-negligible fluctuations and a discrete nature of molecular concentrations, which may in turn alter the frequencies of each chemical reaction event;
indeed, several salient phenomena induced by the smallness in molecule numbers have been studied, both theoretically ~\cite{togashi2001transitions,togashi2003alteration,ohkubo2008transition,biancalani2012noise,biancalani2014noise,awazu2009self,
awazu2007discreteness,samoilov2005stochastic,remondini2013analysis,kobayashi2011connection,shnerb2000importance,togashi2004molecular, 
butler2011fluctuation,saito2015theoretical} and experimentally ~\cite{altschuler2008spontaneous,ma2012small,roostalu2011directional}.
Moreover, how such microscopic molecular discreteness can contribute to cellular functions at a larger scale has gathered much interest, and the effect induced by the discreteness is expected to provide a novel concept to understand cellular behaviors and function~\cite{koumura2014stochasticity,michaelson1999role,ramaswamy2012discreteness,ma2012small,altschuler2008spontaneous}.

In a macroscopic system, i.e., when the volume size of a system and  the number of contained molecules is large, the overall behaviors of the system can be described by
the deterministic rate equation of reaction dynamics for the average concentration of chemicals, or a Langevin equation that takes into account small Gaussian fluctuation around it.

However, the above description is broken down for a small-volume system that contains  a small number of molecules accordingly.
Specifically, several recent theoretical studies have reported that, in certain chemical reaction systems, the chemical compositions and underlying dynamics can be drastically altered under a small-number condition~\cite{togashi2001transitions,togashi2003alteration,ohkubo2008transition,biancalani2012noise,biancalani2014noise,awazu2009self,
awazu2007discreteness,samoilov2005stochastic,remondini2013analysis,kobayashi2011connection,shnerb2000importance,togashi2004molecular, 
butler2011fluctuation,saito2015theoretical,haruna2015distinguishing,matsubara2015optimal} compared to those expected in the rate equation assuming with a large volume and a large number of molecules.
These phenomena are induced by molecular discreteness and its associated stochasticity, designated as ``small-number effects'' or ``discreteness-induced transitions'', which are exemplified by the emergence of multi-modality~\cite{togashi2001transitions,togashi2003alteration,ohkubo2008transition,biancalani2012noise,biancalani2014noise,samoilov2005stochastic,remondini2013analysis,kobayashi2011connection}, reversal of reaction current~\cite{saito2015theoretical} and slow relaxation~\cite{awazu2010discreteness}.

In spite of the several examples of small-number effects reported to date, there is no unified understanding of these phenomena. Indeed, there is no precise definition for these effects; moreover, neither the specific criterion for the ``small-number'' nor the condition for the chemical reaction networks that can generate such effects has been clarified to date.
This situation is due to the lack of a theoretical tool capable of distinguishing between the salient phenomena induced by molecular discreteness and those induced by a trivial discreteness effect.
Here, the term ``trivial discreteness effect'' is used for describing the trivial discrepancy between a continuous and discrete value;
the discrete number divided by the volume does not exactly match with the continuous concentration value in the rate equation. 
In general, any chemical reaction system can show such deviation from the rate equation
and specifically the effect is apparent for a system at a microscopic scale with an extremely small number of molecule (0,1,2,...).
Therefore, it naturally raises the following question: what is the difference between a trivial discreteness effect and the previously reported small-number effect?

In the present paper, we consider the discreteness-induced phenomena that can emerge at a ``mesoscopic'' level, which is clearly larger than the microscopic level (i.e., the scale at which the trivial discreteness effect occurs) but smaller than the macroscopic level, because at the mesoscopic level, the trivial discreteness effect vanishes and only salient phenomena that are relevant at a continuous concentration level remain.
For this purpose, different volume sizes with accordingly different numbers of contained molecules are considered for each chemical reaction system.
The mesoscopic phenomena can be prominent even in the situation where hundreds of molecules exist, which is clearly out of the discrete region; thus, it is reasonable to expect that these phenomena may be relevant in various cellular situations such as the chemical reactions occurring in fine intra-cellular structures.
After introducing the concept of the mesoscopic scale, we propose a criterion for determining whether or not the discreteness-induced phenomena can emerge at the mesoscopic level, and characterize such phenomena occurring at the mesoscopic level as the small-number effect. 
Thus, this criterion enables prediction of whether the small-number effect will emerge in a given chemical reaction network. 
Rather than applying the proposed criterion to one specific system, we here apply it to a general class of chemical reaction systems, and systematically provide a list of the network motifs of chemical reactions that are expected to show the small-number effect.
Furthermore, through motif analysis, we confirm that a previously reported system~\cite{ohkubo2008transition,biancalani2012noise,biancalani2014noise}
 also falls into the motif that is predicted to show the mesoscopic small-number effect.
Specifically, this analysis revealed that a motif involving an autocatalytic reaction (positive feedback) can potentially show the small-number effect, whereas one with an auto-repressive reaction (negative feedback) tends to weaken the small-number effect.
Each predicted small-number effect was confirmed with numerical simulations.

\section{Method: Formulation of Small-Number Effects}
To characterize small-number effects in a chemical reaction system, we begin by focusing on the stationary distribution of the system. Here, we consider only a system under a well-mixed condition, by ignoring spatial inhomogeneity.
Taking the concept of a ‘small number’ literally, the effect we are addressing is not expected to emerge in a “macroscopic” system, and thus should vanish as the number of molecules in the system approaches infinity. 
In addition, we are not interested in a truly ``microscopic'' phenomenon that would emerge under a situation with an extremely low number of molecules, i.e., $1-10$, in which discreteness in the number is dominant. 
Here, we specifically address the phenomenon at a ``mesoscopic'' level, in which stochasticity and molecular discreteness at the ``microscopic'' level exert an influence on a larger scale, but the effect disappears at a macroscopic level. Typically, this effect would be observed at a scale larger than dozens of molecules, and much smaller than the Avogadro number.

We postulate the following two requirements for a chemical reaction system satisfying the conditions of interest.
(i) The ergodicity condition is satisfied and there exists a unique stationary state; hence, divergence in the molecular concentration or the existence of absorbing states is excluded. 
(ii) The discrete probability distribution function $P(\ve{n})$ of the number of each chemical species $\ve{n}=(n_{1},n_{2},...)$ is well approximated by the continuous probability distribution function $P(\ve{x})$ of the concentrations $\ve{x}=(x_{1},x_{2},...)$, in which the concentration of each chemical $x_{i}$ is defined by $x_{i}=n_{i}/\Omega$, according to the volume of the system $\Omega$. 
More concretely, we assume that the time evolution of the probability function is well described by the chemical Fokker-Planck equation~\cite{gillespie2000chemical}, which is derived from the master equation. 

From the requirement (ii), we start from the chemical Fokker-Planck equation. 
The stationary distribution $P_{st}(\ve{x})$ is expected to have the following expression: 
\begin{eqnarray}\label{eq:Pst}
P_{st}(\ve{x})\propto e^{\Omega \phi_{1}(\ve{x})+\phi_{0}(\ve{x})},
\end{eqnarray}
where $\Omega \phi_{1}(\ve{x})$ is the leading order term and $\phi_{0}(\ve{x})$ is the remnant term.
This indicates that $P_{st}(\ve{x})$ approaches $P_{st}(\ve{x}) \propto e^{\Omega \phi_{1}}$ for an infinitely large volume $\Omega \to \infty$. Consequently, $P_{st}(\ve{x})$ has peaks at maxima of $\phi_{1}(\ve{x})$. 
Correspondingly, for this limit, $\ve{x}$ obeys the rate equation, the information of which is contained only in $\phi_{1}(\ve{x})$.
Asymptotically in the limit, small Gaussian fluctuations of the order $1/\sqrt{\Omega}$ exist around the maxima, as demonstrated by van Kampen’s  $\Omega$ expansion~\cite{van1992stochastic}. 
This behavior of $\ve{x}$ at the limit and its asymptotic behaviors are regarded as ``macroscopic'' phenomena.

On the other hand, for small $\Omega$, $\phi_{0}(\ve{x})$ could be of a comparable or dominant order compared to $\Omega \phi_{1}(\ve{x})$ for a certain range 
%$\Delta$ 
of $\ve{x}$. This may result in a drastic change in the distribution, where $P_{st}(\ve{x}) \propto e^{\phi_{0}(\ve{x})}$ is distinguishable from the behavior for $\Omega \to \infty$; this leads to  a small-number effect. The definition of the drastic change will be given below, following derivation of the specific forms of $\phi_{0}$ and $\phi_{1}$.

To define the criterion for the small-number effect, for simplicity, we consider the case of a one-dimensional system that can be described by a one-step Markov process. Extension to multi-dimensional systems is rather straightforward if the detailed balance condition is assumed in addition to assumptions (i) and (ii) above (see Supplemental Materials).
The time evolution of the probability $P(n)$ of the number $n$ of a chemical is given by the following master equation:
\begin{eqnarray}
\frac{\partial P(n)}{\partial t}=T_{n-1}^{+}P(n-1)+T_{n+1}^{-}P(n+1)-(T_{n}^{+}+T_{n}^{-})P(n), \nonumber
\end{eqnarray}
where $T_{n}^{+}$ and $T_{n}^{-}$ are the transition probabilities for $n \to n+1$ and $n \to n-1$, respectively.
Applying the Kramers-Moyal expansion and ignoring orders higher than $1/\Omega^{2}$, the chemical Fokker-Planck equation for the probability $P(x)$ of concentration $x=n/\Omega$ is obtained as follows:
\begin{eqnarray}\label{eq:cFP}
\frac{\partial P(x)}{\partial t}=-\frac{\partial}{\partial x }M_{1}(x)P(x)+\frac{1}{2}\frac{\partial^{2}}{\partial x^{2} }M_{2}(x)P(x),
\end{eqnarray}
in which $\Omega \gg 1$, or at least $\Omega > 1$, is assumed from requirement (ii).
In Eq.~(\ref{eq:cFP}), $M_{1}(x)$ and $M_{2}(x)$ are the first and second moments of the transition probability, respectively, and can be expanded as
\begin{eqnarray}
M_{1}(x)&=&\frac{T^{+}(n)-T^{-}(n)}{\Omega}= m_{1}^{(0)}(x)+\frac{m_{1}^{(1)}(x)}{\Omega} \label{eq:moment1}\\
M_{2}(x)&=&\frac{T^{+}(n)+T^{-}(n)}{\Omega^{2}}=\frac{m_{2}^{(1)} (x)}{\Omega},
\end{eqnarray}
where $m^{(0)}_{i}$ and $m^{(1)}_{i}$ indicate a coefficient of $\Omega^{0}$ and $\Omega^{-1}$ term with $x=n/\Omega$, respectively.
The $\Omega^{-1}$ order term in Eq.~(\ref{eq:moment1}) results from reactions with a reaction order higher than two (e.g., $n(n-1)/\Omega$ in $T^{+}(n)$ or $T^{-}(n)$).
From the boundary condition for the reflecting wall $S(x,t)=M_1(x)P(x,t)-\frac{1}{2}\frac{\partial }{\partial x}M_2(x)P(x,t)=0$ at $x=0$, which is derived from the postulate that $n$ cannot be negative,
the stationary distribution $P_{st}(x)$ of Eq.~(\ref{eq:cFP}) is given by
\begin{eqnarray}
P_{st}(x)&=&\exp \left[ 2\int \frac{M_{1}}{M_{2}}dx' -\log M_{2}+\log C  
\right], \nonumber
\end{eqnarray}
where $C$ is a normalization constant. 
Defining 
\begin{eqnarray}
\phi_{1}(x)=2\int \frac{m_{1}^{(0)}}{m_{2}^{(1)}} dx, \ \ \phi_{0}(x)=2\int \frac{m_{1}^{(1)}}{m_{2}^{(1)}} dx-\log m_{2}^{(1)}, \nonumber
\end{eqnarray}
$P_{st}(x)$ can be obtained as a form of Eq.(\ref{eq:Pst}). 
The derivatives of $\Omega \phi_{1}(x)$ and $\phi_{0}(x)$ are given as
\begin{eqnarray}
\Omega\frac{\partial \phi_{1}}{\partial x}=2\Omega\frac{m_{1}^{(0)}}{m_{2}^{(1)}}, \ \ \ \ \frac{\partial \phi_{0}}{\partial x}=2\frac{m_{1}^{(1)}}{m_{2}^{(1)}}-\frac{1}{m_{2}^{(1)}}\frac{\partial m_{2}^{(1)}}{\partial x}.
\end{eqnarray}
The corresponding rate equation is then written as $\dot{x}=m_{1}^{(0)}$.

Here, we define $\Delta$ as the range within which the inequality
\begin{eqnarray}\label{eq:ast}
\Omega|\phi_{1}'(x)|<|\phi_{0}'(x)|
\end{eqnarray}
is satisfied; i.e., in this range, $\phi_0(x)$ is dominant over $\phi_1(x)$.
Obviously, $\Delta$ cannot be a macroscopic scale, since $\phi_{0}(x)$ cannot be dominant for an infinitely large $\Omega$.
It is important to note that since the concentration $x_{i}$ is defined by $x_{i}=n_{i}/\Omega$, and the genuine distribution $P_{st}(\ve{n})$ is discrete, the distribution $P_{st}(\ve{x})$ in Eq.(\ref{eq:Pst}) is invalid for a scale of $x$ smaller than $\delta x \sim 1/\Omega$. Thus, the relevant resolution of $x_{i}$ in Eq.(\ref{eq:Pst}) is nearly $1/\Omega$.
Hence, if $\Delta$ is smaller than $1/\Omega$, this effect cannot be observed within a meaningful range (e.g., in a stochastic simulation by the Gillespie algorithm~\cite{gillespie1977exact}), and the effect is therefore regarded as representing a ``microscopic'' phenomenon.
If $\Delta \sim 1/\Omega$ (i.e., the range $\Delta$ is in the order of several number of molecules), the effect is also regarded as representing a ``microscopic'' phenomenon.
If $|\Delta| \gg 1/\Omega$, we define the emergence of the small-number effect within $\Delta$ as a ``mesoscopic'' phenomenon.
Specifically, in the following section, 
we assume that the system can show the small-number effect if $|\Delta|$ can be arbitrarily increased by tuning of the kinetic constants or $\Omega$. Note that we assume that the reaction order cannot be arbitrarily large, because many-body reaction higher than four is quite unrealistic.

The peaks of $P_{st}(\ve{x})$ can be shifted owing to the small-number effect. A peak for $\Omega \to \infty$ is given by $x_{1}$, which is a stable fixed point of the rate equation (i.e., $0= m_1^{(0)}(x_{1})=\partial \phi_1/\partial x |_{x=x_1}$), whereas that for finite $\Omega$ is given by $x_{p}$, which is a solution of $0=\Omega\partial \phi_{1}/\partial x+\partial \phi_{0}/\partial x$. 
Only when $|x_{p}-x_{1}| \gg 1/\Omega$ (i.e., $|x_{p}-x_{1}|$ can be arbitrarily larger than $1/\Omega$), we consider that the observed shift represents a ``mesoscopic phenomenon''. 

\section{Results}
\subsection{Systematic Listing of Reaction Motifs}
Using the criterion proposed above, we can now determine whether or not a given chemical reaction system will exhibit the small-number effect. 
Here, instead of focusing only on one specific chemical reaction system, we address a more general question: what type of network topologies can be expected to show small-number effects? To answer this question, we consider all possible chemical reaction systems consisting of
one or two chemical species, and for each topology, we examine whether or not the small-number effects can emerge
according to the proposed criterion.

\begin{figure}[h]
\includegraphics[width=7cm]{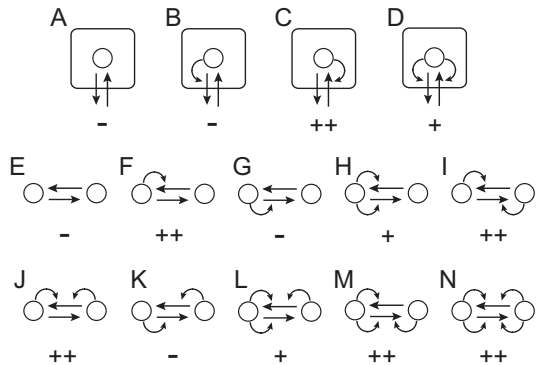}
\caption{
The listed motifs consisting of one or two chemical species.
The ``++'' symbol indicates a motif that can show a small-number effect at the lowest reaction order, whereas a motif represented by ``+'' can show a small-number effect at a higher order.
The ``-'' symbol indicates a motif that cannot show the small-number effect for any choice of kinetic constant.
Detailed information on the behavior and analytic expressions of $\phi_{1}'$ and $\phi_{0}'$ for each motif in the list is given in the Supplemental Materials.
}
\label{fig:list}
\end{figure}
The resulting motif list is shown in Fig.~\ref{fig:list}. 
The network motifs A-D represent chemical reaction systems with one chemical species; an arrow pointing out of (into) the box represents annihilation (creation) of a molecule, whereas an arrow pointing to another arrow indicates that the molecule also acts as a catalyst of the reaction. 
The motifs E-N represent chemical reaction systems with two chemical species, in which the total number of molecules is conserved;
an arrow connecting one circle to another circle indicates a substrate-product relationship, whereas an arrow connecting a circle to another arrow indicates that the reaction is catalyzed by the molecule represented by the circle.
Note that for a motif with an autocatalytic reaction (e.g., C, D, or F), the reactions will stop once the number of autocatalytic molecules becomes zero, and thus such a motif cannot satisfy the ergodicity condition (i). To prevent this ergodicity-breaking, the creation of molecule $ \emptyset \to X$ was added to motifs C and D, whereas $X \leftarrow Y$ or $X \leftrightarrow Y$ was added to the motifs F, H, I, J, L, M, and N.

Analysis for each motif is demonstrated in the following section and in Supplemental Materials.
The ``++'' symbol in Fig.~\ref{fig:list} indicates a motif that can potentially show a small-number effect at the lowest reaction order, whereas a ``+'' motif cannot show a small-number effect at the lowest reaction order, but can show such effects at a higher order. 
For instance, motif D, which consists of $\emptyset \to X$, $lX\to (l+1)X$, and $mX\to (m-1)X$ with integer reaction orders $l\ge 1$ and $m\ge 2$, is marked as ``+'' in Fig.~\ref{fig:list}, because the motif cannot show the small-number effect for the lowest reaction order $l=1$, $m=2$ but can show the effect for $l=m=2$.
The ``-'' symbol indicates a motif that cannot show the small-number effect for any choice of kinetic constant or volume size, unless the reaction order is unrealistically high.

\subsection{Representative Examples}
We here provide detailed examples of motif analysis for three representative motifs of a chemical reaction system; analysis of all other motifs is presented in the Supplemental Materials.
The first example does not show the small-number effect for any choice of parameters, the second example shows a power-law tail distribution as well as a shift in the peak position due to the small-number effect, and the third example shows the emergence of multi-modality, as reported in \cite{ohkubo2008transition,biancalani2012noise,biancalani2014noise}.

\subsubsection{Motif A}
The first example is motif A in Fig.\ref{fig:list}, which consists of one chemical species, X, and the following two chemical reactions:
\begin{eqnarray}
&X\xrightarrow[1]{} \emptyset, & \ \ \  \emptyset \xrightarrow[d]{} X.
\end{eqnarray}	
By denoting $n$ as the number of $X$ molecules, the transition probabilities of $n \to n+1$ and $n \to n-1$ are given as
\begin{eqnarray}
&T^{+}_{n}&=d\Omega, \ \ T^{-}_{n}=n.
\end{eqnarray}
Then, the corresponding chemical Fokker-Planck equation is obtained by
\begin{equation}
\frac{\partial P(x,t)}{\partial t} =-\frac{\partial}{\partial x}(d-x)P(x,t)+\frac{1}{2\Omega}\frac{\partial^{2}}{\partial x^{2}}(d+x)P(x,t),
\end{equation}
which gives
\begin{eqnarray}
\phi_{1}'&=& 2\frac{d-x}{d+x}, \ \ \ \phi_{0}'=-\frac{1}{(d+x)}.
\end{eqnarray}
The condition Eq.~(\ref{eq:ast}) is then written as $|d-x| <1/2\Omega$, implying that the range $\Delta$, satisfying Eq.~(\ref{eq:ast}), is narrower than $1/\Omega$.
Therefore, $\phi_{0}(x)$ cannot be dominant for any choice of $d$ and $\Omega$, and thus the small-number effect cannot emerge in this motif.

The shift of the peak due to the contribution of $\phi_{0} (x)$ is calculated as $x_{1}-x_{p}=1/2\Omega$, indicating that the magnitude of the shift is less than one molecule, which is consistent with our claim that this motif does not show the small-number effect.

\subsubsection{Motif C}
\begin{figure}[h]
\hspace{-0.5cm}
\includegraphics[width=9cm]{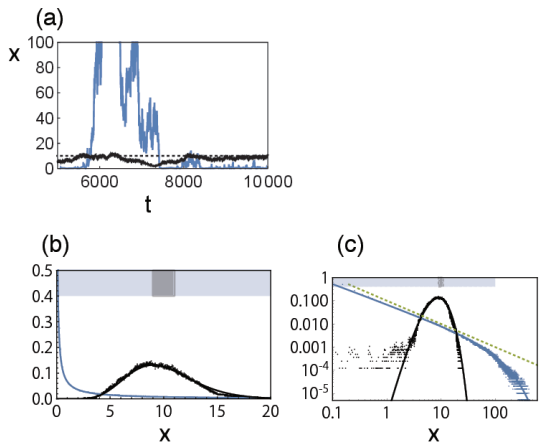}
\caption{The profiles of the chemical concentration of molecule X in motif C. The parameters k = 0.999 and d = 0.01 were used.
In each graph, black indicates the result for a large volume size ($\Omega=1000$), whereas blue indicates the result for a small volume size ($\Omega=10$).
(a) The time series of the concentration $x$. The black broken line shows the concentration at the fixed point of the rate equation. 
(b) The stationary distribution of the concentration. The lines indicate the theoretical estimates $P \propto \exp [ \Omega \phi_{1}+\phi_{0}]$, and the dots indicate the results of the simulation with the Gillespie algorithm. The color band at the top of the graph represents the range $\Delta$, where the small-number effect emerges for each $\Omega$.
(c) The log-log plot of (b). The dotted green line represents $x^{-1}$ for reference.
}\label{fig:for_motifC}
\end{figure}

Next, we consider the case of motif C, which is obtained by addition of the positive feedback reaction $l X\to (l+1)X$ to motif A, where $l$ is a positive integer. 
Note that the reaction $\emptyset \to X$ is necessary because a system without this reaction has an absorbing state $n=0$, which is not allowed according to requirement (i).
This motif can show the small-number effect even for the lowest reaction order $l=1$.
For the case of $l=1$, the transition probability is given as
\begin{eqnarray}
&T^{+}_{n}&=d\Omega +kn \ \ \ T^{-}_{n}=n,
\end{eqnarray}
where $k$ is the rate constant of the positive feedback, which satisfies $k<1$ from requirement (i), since $x=n/\Omega$ diverges at the rate equation for $k >1$. The corresponding $\phi_{1}'(x)$ and $\phi_{0}'(x)$ are calculated as
\begin{equation}
\phi'_1(x)=2 \frac{d+(k-1)x}{d+(k+1)x},  \ \ \ \phi'_0(x)=-\frac{1+k}{d+(1+k)x}.
\end{equation}
Denoting the peak of $\phi_{1}(x)$ by $x_{1}=d/(1-k)$, the condition for Eq.~(\ref{eq:ast}) is written as
\begin{eqnarray}
|x-x_1|<\frac{1+k}{1-k}\frac{1}{2\Omega},
\end{eqnarray}
implying that $\Delta$ can be arbitrarily larger than $1/\Omega$ if $k$ is sufficiently close to  unity.
Therefore, this motif can show the small-number effect as the crossover from $P_{st} \propto \exp (\Omega \phi_{1}(x))$ to $P_{st} \propto \exp (\phi_{0}(x))$, where
\begin{eqnarray}
 e^{\Omega \phi_{1}}&=&\{ d+(1+k)x \}^{\frac{4\Omega d}{(1+k)^{2}}}e^{-2\Omega \frac{1-k}{1+k}x} \\
 e^{ \phi_{0}} &=&\{ d+(1+k)x \}^{-1}.
\end{eqnarray}
This small-number effect is observed as the emergence of a power-law distribution within $\Delta$, as shown in Fig.\ref{fig:for_motifC}(b) and (c). 
Temporally intermittent bursting of $x$ is also observed in concert with the power-law tail, which is depicted in Fig.\ref{fig:for_motifC}(a).
The peak of $P_{st}(x)$ is shifted to the left due to the contribution of $\phi_{0}(x)$, and the shift is estimated as $(1+k)/2\Omega(1-k)$, which is also significant for $k \to 1$.
This emergence of the power-law tail distribution due to the small-number effect in a simple reaction system, which has not been reported to date, provides a clear example that cannot be explained by simple fluctuation around a fixed point.
Note that, for building histogram using Gillespie algorithm, samples should be recorded at each regular time interval (e.g. $dt=0.05$), rather than sampling synchronously with the algorithm step (e.g. sampling each 10 algorithm step), otherwise a large sampling bias may be introduced.

\subsubsection{Motif  I}
\begin{figure}[h]
 \includegraphics[width=7.5cm]{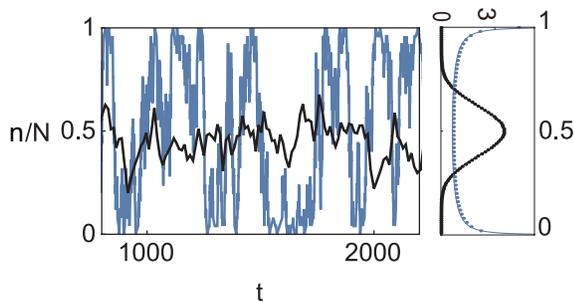}
\caption{Profiles of the chemical concentration of molecule X in motif I, for $\mu=0.01$.
The left  panel illustrates the time series of $x=n/N$, and the right panel shows the probability distribution of the concentration.
The blue line indicates the result for a small number of molecules ($N=50$) and the black line represents the result for a large number of molecules ($N=1000$). The lines in the right panel indicate the theoretical estimates of $P \propto \exp [ \Omega \phi_{1}+\phi_{0}]$, and the dots indicate the results of the simulation using the Gillespie algorithm. Analytical and numerical results are overlapped completely.}
\label{fig:for_motif_I}
\end{figure}

As the final example, we consider motif I, consisting of two chemical species, X, Y, and the following four reactions:
\begin{eqnarray}
Y \xrightarrow[\mu_{1}]{} X, \ \  X \xrightarrow[\mu_{2}]{} Y, \ \   \ X + Y\xrightarrow[k_{1}]{} 2X, \  \  \ X+Y \xrightarrow[k_{2}]{} 2Y, \nonumber
\end{eqnarray}; this type of motif has previously been shown to exhibit the small-number effect~\cite{ohkubo2008transition,biancalani2012noise,biancalani2014noise}.
Note that the first two reactions are necessary to satisfy the ergodicity condition, because the absence of either reaction gives rise to an absorbing state.
Since the total number of molecules $N$ is conserved throughout the above reactions, $N$ is set to be identical to a unitless volume $\Omega$ via proper transformation.
By denoting $n$ as the number of $X$ molecules and $N-n$ as the number of $Y$ molecules, the transition probabilities of $n \to n+1$ and $n \to n-1$ are given as
\begin{eqnarray}
T^{+}_{n}&=&\mu_{1}(N-n)+k_{1}n(N-n)/N, \nonumber \\ 
T^{-}_{n}&=&\mu_{2}n+k_{2}n(N-n)/N.
\end{eqnarray}
The corresponding $\phi_{1}'(x)$ and $\phi_{0}'(x)$ are calculated as
 \begin{eqnarray}
 \phi'_1(x)&=&2\frac{(k_{2}-k_{1})x^{2}-(k_{2}-k_{1}+\mu_{1}+\mu_{2})x+\mu_{1}}{\mu_{1}(1-x)+\mu_{2}x+(k_{1}+k_{2})x(1-x)},  \nonumber \\ 
 \phi'_0(x)&=&\frac{2(k_{1}+k_{2})x-(k_{1}+k_{2}-\mu_{1}+\mu_{2})}{\mu_{1}(1-x)+\mu_{2}x+(k_{1}+k_{2})x(1-x)}.
\end{eqnarray}
Then, Eq.~(\ref{eq:ast}) is written as
\begin{eqnarray}
|(k_{2}-k_{1})x^{2}-(k_{2}-k_{1}+\mu_{1}+\mu_{2})x+\mu_{1}|  \nonumber\\
<\epsilon |2(k_{1}+k_{2})x-(k_{1}+k_{2}-\mu_{1}+\mu_{2})|,
\end{eqnarray}
where $\epsilon=1/2N$.
For simplicity, we focus on the case with $k_{1}=k_{2}=1$ and $\mu_{1}=\mu_{2}=\mu$,
from which Eq.~(\ref{eq:ast}) is rewritten as $\epsilon > \mu/2$.
This implies that $\Delta=[0,1]$ for $N<1/\mu$, and $\Delta=\emptyset$ otherwise. Specifically,
for $\mu \ll 1$ and $N \sim 1/\mu$, $|\Delta|$ is sufficiently larger than $1/N$ because $|\Delta|=1 \gg 1/N$; thus, this motif can show the small-number effect.
In this case, $e^{N\phi_{1}(x)}$ and $e^{\phi_{0}(x)}$ are calculated as
\begin{eqnarray}
 e^{N\phi_{1}}&=&\{ -x^{2}+x+\mu /2\}^{\mu N } \\
 e^{ \phi_{0}} &=&\{ -x^{2}+x+\mu /2\}^{-1},
\end{eqnarray}
where the former distribution is unimodal, whereas the latter is bimodal. At the critical value $N=1/\mu$, $N\phi_{1}+\phi_{0}$ is zero for the entire region of $x$, and thus the distribution is uniform.
Thus, by decreasing the total number of molecules, bimodality in the distribution function $P_{st}(x)$ emerges, and switching behavior between $x=0$ and $1$ correspondingly appears in the time series, as shown in Fig.~\ref{fig:for_motif_I}.
%Our analysis clearly shows that this emergence of bistability is not real phase transition in a sense of statistical physics, but a crossover between $\phi_{1}(x)$ and $\phi_{0}(x)$.

 %%%REVISED%%

 \section{Discussion}
As confirmed numerically, our proposed criterion could successfully predict the small-number effect at a ``mesoscopic'' scale.
The representative examples outlined above show not only the emergence of multi-modality (e.g. in motif I) but also the emergence of a power-law tail distribution (and corresponding bursting behavior in the time series) in motif C, as illustrated in Fig.~\ref{fig:for_motifC}.
Although the emergence of the power-law tail due to self-organized criticality has been reported previously~\cite{awazu2009self,
awazu2007discreteness} using complex chemical reaction networks, our present finding does not rely on this mechanism. We consider that motif C can serve as a minimal model to exhibit the emergence of the power law due to the small-number effect.

Although we adopted the chemical Fokker-Planck equation, which is relevant to continuous variables, we also incorporated information of the discrete nature of molecule numbers as a resolution in the variables, which resulted in the criterion $|\Delta| \gg 1/\Omega$. 
With this criterion, microscopic phenomena are successfully excluded from the small-number effect. For instance, the shift of the peak in motif A, which was predicted by the chemical Fokker-Planck equation, was concluded to not be relevant at a mesoscopic scale.
This demonstrates that a given phenomenon described in the chemical Fokker-Planck equation does not always indicate a mesoscopic phenomenon, and that phenomenon occurring at every scale can be included in the equation.
Therefore, the proposed treatment could successfully combine the analysis of a continuous equation for molecular concentration with the information of discreteness in molecule number.
From this aspect, our study is different from previous works on noise-induced transitions~\cite{suzuki1981phase,horsthemke1984noise}, which focus on the impact of the noise effect on the distribution of the Langevin equation. In these previous studies, the elemental process behind the Langevin equation was ignored, and thereby information of the molecular discreteness was also disregarded; thus, the microscopic and mesoscopic phenomena could not be distinguished. 
For example, the noise-induced transition framework predicts the shift of the peak for all the motifs in Fig. 1, whereas for motifs classified as (-) the shift cannot be observed at a mesoscopic scale.

Note that the derivation of chemical Fokker-Planck is not  ``systematic'' in the sense of perturbation theory, which is in contrast to the linear noise approximation (LNA) derived by the van Kampen's expansion~\cite{van1992stochastic}.
Here we adopted chemical Fokker-Planck equation, that is also derived from the chemical Langevin equation\cite{gillespie2000chemical}. As is discussed in introduction, recent numerical studies support the validity of chemical Fokker-Planck equation up to smaller value of $\Omega$ than the range adopted for LNA\cite{biancalani2014noise,mckane2014stochastic,grima2011accurate}. In fact, we also demonstrated that the direct numerical simulations agree quite well with the predictions of chemical Fokker-Planck equation for all motifs marked by ``+'' or ``++'' in Fig.1, where small-number effects are well reproduced by the chemical Fokker-Planck equation, rather than by the LNA.

The list of networks shown in Fig.~\ref {fig:list} indicates that motifs showing the small-number effect always involve an autocatalytic reaction (positive feedback), which in turn suggests that autocatalysis is an essential structure for the emergence of the small-number effect. In addition, comparison of motif C with D or motif J with L suggests that an auto-repressive reaction (negative feedback) tends to prevent emergence of the small-number effect.  
Indeed, most of the previous studies that reported small-number effects considered chemical reactions containing autocatalysis~\cite{togashi2001transitions,togashi2003alteration,ohkubo2008transition,biancalani2012noise,biancalani2014noise,kobayashi2011connection,shnerb2000importance,togashi2004molecular, butler2011fluctuation,saito2015theoretical}.

The small-number effect illustrated here demonstrates that the number of molecules, rather than the concentration, can drastically alter the stationary state of the system.
Moreover, the reversal of the current of a chemical reaction has been reported~\cite{saito2015theoretical} due to the small-number effect.
These examples suggest a novel way of regulating a chemical reaction, that is, regulation based on the number of molecules (or equivalently volume size).
This number-based regulation can be further extended to consider the effective number of molecules, rather than the total number of molecules, where ‘effective’ means the number of molecules that can join together in a chemical reaction under a restricted condition.  For example, under a molecular crowding condition, each molecule is accessible to only  a small number of surrounding molecules, which may induce the small-number effect, as reported here. 

 \section{Summary}
In this paper, we define the small-number effect of a well-stirred system as a mesoscopic phenomenon by focusing on the stationary distribution, and propose a criterion for determining whether or not this effect can emerge in a given chemical reaction network. 
By examining all possible chemical reaction networks consisting of one or two chemical species, we have provided a list of the network motifs of a chemical reaction that can potentially show the small-number effect.
The motifs considered herein are quite simple and can thus easily be incorporated as a subpart for a complex, real biochemical reaction network.
%The motifs considered here are quite simple so that they are easily included in a complex, real biochemical reaction network
 For example, similar biochemical reactions to motif I have been studied in several biological contexts~\cite{artyomov2007purely, altschuler2008spontaneous,kobayashi2011connection,jafarpour2015noise,dyson2015onset}.
The list of motifs provided herein should be helpful in searching for other candidates of biochemical reactions in which the small-number effect is expected to play an important role within a cell. 
Furthermore, considering recent advances in synthetic biology~\cite{lee1996self,soloveichik2010dna,padirac2012bottom}, the list is also expected to be useful for designing a system that shows the small-number effect. 
Of course, our criterion can also be applied to larger network motifs using the criterion described for a multivariable system given in the Supplemental Materials.
Extensions of our theory to reaction-diffusion systems will be an important issue in the future.

\section{supplementary material}
See supplementary material for the extension of the proposed theory to the multivariate system, and the details of the motif analysis.

\begin{acknowledgments}
This work was supported by a Grant-in-Aid for Scientific Research on Innovative Areas: ``Spying minority in biological phenomena (No. 3306)" (26115704), and by the Platform Project for Supporting in Drug
Discovery and Life Science Research (Platform for Dynamic
Approaches to Living System) from Japan Agency
for Medical Research and Development(AMED).
\end{acknowledgments}

\bibliographystyle{unsrt}
%\bibliography{saitoref_4}

\begin{thebibliography}{10}

\bibitem{taniguchi2010quantifying}
Yuichi Taniguchi, Paul~J Choi, Gene-Wei Li, Huiyi Chen, Mohan Babu, Jeremy
  Hearn, Andrew Emili, and X~Sunney Xie.
\newblock Quantifying e. coli proteome and transcriptome with single-molecule
  sensitivity in single cells.
\newblock {\em Science}, 329(5991):533--538, 2010.

\bibitem{ishihama2008protein}
Yasushi Ishihama, Thorsten Schmidt, Juri Rappsilber, Matthias Mann, F~Ulrich
  Hartl, Michael~J Kerner, and Dmitrij Frishman.
\newblock Protein abundance profiling of the escherichia coli cytosol.
\newblock {\em BMC genomics}, 9(1):102, 2008.

\bibitem{togashi2001transitions}
Yuichi Togashi and Kunihiko Kaneko.
\newblock Transitions induced by the discreteness of molecules in a small
  autocatalytic system.
\newblock {\em Physical review letters}, 86(11):2459, 2001.

\bibitem{togashi2003alteration}
Yuichi Togashi and Kunihiko Kaneko.
\newblock Alteration of chemical concentrations through discreteness-induced
  transitions in small autocatalytic systems.
\newblock {\em Journal of the Physical Society of Japan}, 72(1):62--68, 2003.

\bibitem{ohkubo2008transition}
Jun Ohkubo, Nadav Shnerb, and David A.~Kessler.
\newblock Transition phenomena induced by internal noise and quasi-absorbing
  state.
\newblock {\em Journal of the Physical Society of Japan}, 77(4), 2008.

\bibitem{biancalani2012noise}
Tommaso Biancalani, Tim Rogers, and Alan~J McKane.
\newblock Noise-induced metastability in biochemical networks.
\newblock {\em Physical Review E}, 86(1):010106, 2012.

\bibitem{biancalani2014noise}
Tommaso Biancalani, Louise Dyson, and Alan~J McKane.
\newblock Noise-induced bistable states and their mean switching time in
  foraging colonies.
\newblock {\em Physical Review Letters}, 112(3):038101, 2014.

\bibitem{awazu2009self}
Akinori Awazu and Kunihiko Kaneko.
\newblock Self-organized criticality of a catalytic reaction network under
  flow.
\newblock {\em Physical Review E}, 80(1):010902, 2009.

\bibitem{awazu2007discreteness}
Akinori Awazu and Kunihiko Kaneko.
\newblock Discreteness-induced transition in catalytic reaction networks.
\newblock {\em Physical Review E}, 76(4):041915, 2007.

\bibitem{samoilov2005stochastic}
Michael Samoilov, Sergey Plyasunov, and Adam~P Arkin.
\newblock Stochastic amplification and signaling in enzymatic futile cycles
  through noise-induced bistability with oscillations.
\newblock {\em Proceedings of the National Academy of Sciences of the United
  States of America}, 102(7):2310--2315, 2005.

\bibitem{remondini2013analysis}
Daniel Remondini, Enrico Giampieri, Armando Bazzani, Gastone Castellani, and
  Amos Maritan.
\newblock Analysis of noise-induced bimodality in a michaelis--menten
  single-step enzymatic cycle.
\newblock {\em Physica A: Statistical Mechanics and its Applications},
  392(2):336--342, 2013.

\bibitem{kobayashi2011connection}
Tetsuya~J Kobayashi.
\newblock Connection between noise-induced symmetry breaking and an
  information-decoding function for intracellular networks.
\newblock {\em Physical review letters}, 106(22):228101, 2011.

\bibitem{shnerb2000importance}
Nadav~M Shnerb, Yoram Louzoun, Eldad Bettelheim, and Sorin Solomon.
\newblock The importance of being discrete: Life always wins on the surface.
\newblock {\em Proceedings of the National Academy of Sciences},
  97(19):10322--10324, 2000.

\bibitem{togashi2004molecular}
Yuichi Togashi and Kunihiko Kaneko.
\newblock Molecular discreteness in reaction-diffusion systems yields steady
  states not seen in the continuum limit.
\newblock {\em Physical Review E}, 70(2):020901, 2004.

\bibitem{butler2011fluctuation}
Thomas Butler and Nigel Goldenfeld.
\newblock Fluctuation-driven turing patterns.
\newblock {\em Physical Review E}, 84(1):011112, 2011.

\bibitem{saito2015theoretical}
Nen Saito and Kunihiko Kaneko.
\newblock Theoretical analysis of discreteness-induced transition in
  autocatalytic reaction dynamics.
\newblock {\em Physical Review E}, 91(2):022707, 2015.

\bibitem{altschuler2008spontaneous}
Steven~J Altschuler, Sigurd~B Angenent, Yanqin Wang, and Lani~F Wu.
\newblock On the spontaneous emergence of cell polarity.
\newblock {\em Nature}, 454(7206):886--889, 2008.

\bibitem{ma2012small}
Rui Ma, Jichao Wang, Zhonghuai Hou, and Haiyan Liu.
\newblock Small-number effects: a third stable state in a genetic bistable
  toggle switch.
\newblock {\em Physical review letters}, 109(24):248107, 2012.

\bibitem{roostalu2011directional}
Johanna Roostalu, Christian Hentrich, Peter Bieling, Ivo~A Telley, Elmar
  Schiebel, and Thomas Surrey.
\newblock Directional switching of the kinesin cin8 through motor coupling.
\newblock {\em Science}, 332(6025):94--99, 2011.

\bibitem{koumura2014stochasticity}
Takuya Koumura, Hidetoshi Urakubo, Kaoru Ohashi, Masashi Fujii, and Shinya
  Kuroda.
\newblock Stochasticity in ca 2+ increase in spines enables robust and
  sensitive information coding.
\newblock {\em PLoS ONE}, 9:e99040, 2014.

\bibitem{michaelson1999role}
JAMES Michaelson.
\newblock The role of molecular discreteness in normal and cancerous growth.
\newblock {\em Anticancer research}, 19(6):4853--4868, 1999.

\bibitem{ramaswamy2012discreteness}
Rajesh Ramaswamy, N{\'e}lido Gonz{\'a}lez-Segredo, Ivo~F Sbalzarini, and Ramon
  Grima.
\newblock Discreteness-induced concentration inversion in mesoscopic chemical
  systems.
\newblock {\em Nature communications}, 3:779, 2012.

\bibitem{haruna2015distinguishing}
Taichi Haruna.
\newblock Distinguishing between discreteness effects in stochastic reaction
  processes.
\newblock {\em Physical Review E}, 91(5):052814, 2015.

\bibitem{matsubara2015optimal}
Yoshiya~J Matsubara and Kunihiko Kaneko.
\newblock Optimal system size for emergence of self-replicating polymer system.
\newblock {\em arXiv preprint arXiv:1509.08865}, 2015.

\bibitem{awazu2010discreteness}
Akinori Awazu and Kunihiko Kaneko.
\newblock Discreteness-induced slow relaxation in reversible catalytic reaction
  networks.
\newblock {\em Physical Review E}, 81(5):051920, 2010.

\bibitem{gillespie2000chemical}
Daniel~T Gillespie.
\newblock The chemical langevin equation.
\newblock {\em The Journal of Chemical Physics}, 113(1):297--306, 2000.

\bibitem{van1992stochastic}
Nicolaas~Godfried Van~Kampen.
\newblock {\em Stochastic processes in physics and chemistry}, volume~1.
\newblock Elsevier, 1992.

\bibitem{gillespie1977exact}
Daniel~T Gillespie.
\newblock Exact stochastic simulation of coupled chemical reactions.
\newblock {\em The journal of physical chemistry}, 81(25):2340--2361, 1977.

\bibitem{suzuki1981phase}
Masuo Suzuki, Kunihiko Kaneko, and Fumiyoshi Sasagawa.
\newblock Phase transition and slowing down in non-equilibrium stochastic
  processes.
\newblock {\em Progress of Theoretical Physics}, 65(3):828--849, 1981.

\bibitem{horsthemke1984noise}
Werner Horsthemke and Ren{\'e} Lefever.
\newblock Noise-induced transitions in physics, chemistry, and biology.
\newblock {\em Noise-Induced Transitions: Theory and Applications in Physics,
  Chemistry, and Biology}, pages 164--200, 1984.

\bibitem{mckane2014stochastic}
Alan~J McKane, Tommaso Biancalani, and Tim Rogers.
\newblock Stochastic pattern formation and spontaneous polarisation: the linear
  noise approximation and beyond.
\newblock {\em Bulletin of mathematical biology}, 76(4):895--921, 2014.

\bibitem{grima2011accurate}
Ramon Grima, Philipp Thomas, and Arthur~V Straube.
\newblock How accurate are the nonlinear chemical fokker-planck and chemical
  langevin equations?
\newblock {\em The Journal of chemical physics}, 135(8):084103, 2011.

\bibitem{artyomov2007purely}
Maxim~N Artyomov, Jayajit Das, Mehran Kardar, and Arup~K Chakraborty.
\newblock Purely stochastic binary decisions in cell signaling models without
  underlying deterministic bistabilities.
\newblock {\em Proceedings of the National Academy of Sciences},
  104(48):18958--18963, 2007.

\bibitem{jafarpour2015noise}
Farshid Jafarpour, Tommaso Biancalani, and Nigel Goldenfeld.
\newblock A noise-induced mechanism for biological homochirality of early life
  self-replicators.
\newblock {\em arXiv preprint arXiv:1507.00044}, 2015.

\bibitem{dyson2015onset}
Louise Dyson, Christian~A Yates, Jerome Buhl, and Alan~J McKane.
\newblock Onset of collective motion in locusts is captured by a minimal model.
\newblock {\em Physical Review E}, 92(5):052708, 2015.

\bibitem{lee1996self}
David~H Lee, Juan~R Granja, Jose~A Martinez, Kay Severin, and M~Reza Ghadiri.
\newblock A self-replicating peptide.
\newblock {\em Nature}, 382(6591):525--528, 1996.

\bibitem{soloveichik2010dna}
David Soloveichik, Georg Seelig, and Erik Winfree.
\newblock Dna as a universal substrate for chemical kinetics.
\newblock {\em Proceedings of the National Academy of Sciences},
  107(12):5393--5398, 2010.

\bibitem{padirac2012bottom}
Adrien Padirac, Teruo Fujii, and Yannick Rondelez.
\newblock Bottom-up construction of in vitro switchable memories.
\newblock {\em Proceedings of the National Academy of Sciences},
  109(47):E3212--E3220, 2012.

\end{thebibliography}

\end{document}